\documentclass[usenatbib]{mnras}

\usepackage{lineno} 
\usepackage{txfonts}
\usepackage{graphicx}

\newcommand{\hess}{\textsc{H.E.S.S.}}

\newcommand{\swiftxrt}{\textsl{Swift}/XRT}

\newcommand{\fer}{{\sl {\it Fermi}}}
\newcommand{\fla}{\fer-LAT}
\newcommand{\rxte}{\textsl{RXTE}}
\newcommand{\xmm}{\textsl{XMM}-Newton}
\newcommand{\pks}{PKS~2155$-$304\ }
\newcommand{\pksc}{PKS~2155$-$304}

\newcommand{\fvar}{F_{\mathrm{var}}}

\newcommand{\gr}{$\gamma$-ray}
\newcommand{\grs}{$\gamma$-rays}

\begin{document}

\title[Variability studies of PKS 2155-304]{Variability studies and modeling of the blazar PKS~2155-304 in the light of a decade of multi-wavelength observations}

\author[J.~Chevalier et al.]{
J.~Chevalier,$^1$ D.A.~Sanchez,$^1$\thanks{david.sanchez@lapp.in2p3.fr} P.~D.~Serpico,$^2$ J.-P.~Lenain,$^3$ G.~Maurin$^1$\\
$^1$Univ. Grenoble Alpes, Univ. Savoie Mont Blanc, CNRS, LAPP, 74000 Annecy, France \\
$^2$Univ. Grenoble Alpes, USMB, CNRS, LAPTh, F-74940 Annecy, France \\
$^3$Sorbonne Universit\'e, Universit\'e Paris Diderot, Sorbonne Paris Cit\'e, CNRS/IN2P3, Laboratoire de Physique Nucl\'eaire et de Hautes Energies, LPNHE,\\ 4 Place Jussieu, F-75252 Paris, France
}

\maketitle

\begin{abstract}The variability of the high-frequency peaked BL Lac object \pks\ is studied using almost 10~years of optical, X-ray and \grs\ data. Publicly available data have been gathered and analyzed with the aim to characterize the variability and to search for log-normality or periodic behavior. The optical and X-ray range follow a log-normal process; a hint for a periodicity of about $\approx$ 700 days is found in optical and in the high energy (100 MeV $<$ E $<$ 300 GeV) range. A one zone, time-dependent, synchrotron self-Compton model is successfully used to reproduce the evolution with energy of the variability and the tentatively reported periodicity.
\end{abstract}



\section{Introduction}

Blazars are a subclass of active galactic nuclei with a relativistic jet pointed towards the Earth. The observed emission, from radio up to TeV energies, is dominated by their jet. Nevertheless, the precise composition of the jet  as well as the acceleration and emission processes involved are not known. If the jet is dominated by leptons ($e^\pm$ pairs), leptonic models such as the synchrotron self-Compton \citep[SSC]{THEO::SSC_BAND} or  the external Compton \citep[]{Dermer1993} are invoked to reproduce the electromagnetic emission. These models differ essentially in the target photon field for the inverse Compton emission. Alternatively, hadronic models have been considered, where \grs\ are emitted through photo-production of pions \citep[e.g. ][]{1993A&A...269...67M} or synchrotron emission of protons \citep[e.g. ][]{2000NewA....5..377A}. Unfortunately, to disentangle between these models, fits of the time-averaged spectral energy distribution (SED) are insufficient.

However, one of the striking properties of blazars is their variability\footnote{Another diagnostic tool is provided by multimessenger studies, notably in high-energy neutrinos, which till now is however limited to the single spectacular case of the blazar TXS 0506+056~\citep[]{IceCube:2018dnn}.}. Their brightness can vary at time scales ranging from minutes to years, and this behavior has been observed at all wavelengths, from radio waves to X-rays and gamma rays. The two classes of models predict different variability patterns.  Hence, long-term observations and statistical studies of the variability are key tools in the comprehension of these objects.

\pks \citep[redshift $z=0.116$; ][]{1993ApJ...411L..63F} is a very well known blazar detected at TeV energies for the first time in 1999 \citep{1999ApJ...513..161C} and later confirmed by \citet{2005A&A...430..865A}. It has been classified as high-frequency peaked (HBL) thanks to X-ray observations from the HEAO-1 satellite \citep{1979ApJ...229L..53S}. At TeV energies, this object exhibits large flares on minutes timescale \citep{2007ApJ...664L..71A} but also variation on longer timescales \citep{2016arXiv161003311H}. At lower energy, \fla\ reported variability on monthly timescale \citep{2015ApJS..218...23A} as well as much more rapid flares \citep{2014ATel.6148....1C,2013ATel.4755....1C}. In X-ray, the source exhibits variability \citep[see e.g. ][]{2015ASInC..12...77G} and even intra-day variability was reported \citep{2017ApJ...841..123P}.

\pks variability in optical and X-ray can be deeply studied thanks to systematic observations by SMARTS, \rxte, \swiftxrt\ and \xmm. With the impressive dataset recorded by \hess\ in the TeV range over 9~years \citep{2016arXiv161003311H} and the constant monitoring of the \fla, such studies can be extended to the \gr\ band as well.

This article is structured as follows:
 Section~\ref{sec:dataset} presents the multiwavelength dataset used. Section~\ref{sec:results} details our analyses on the variability of \pksc. A time-dependent modeling is presented in Section~\ref{sec:sscmodeling}. In Section~\ref{sec:discconcl}, we summarize our results and conclude.


\section{Observations and analysis}
\label{sec:dataset}

\subsection{\grs\ datasets}
In the very high energy (VHE, $E\gtrsim 100$ GeV) range, this work makes use of the \hess\ data presented in \citet{2016arXiv161003311H}, reporting on  9~years of observations of \pksc. In the high energy (HE, 100 MeV $<$ E $\lesssim$ 300 GeV) range, the \fer\ data presented in~\citet[]{2015arXiv150903104C}, where the light curves have been computed in two energy bands, are also used.

The possibility to extend the \fer\ (as well as X-ray, and optical) light curve to further data taking periods was considered since---contrarily to  \hess\ data for an article external to the collaboration as this one---the former are available. Nevertheless, we deemed very important  the role of the \hess\ instrument for the phenomenological interpretation, since probing a unique spectral range, and thus more helpful than the added value of a data taking extension limited to the lower energies.

\subsection{X-ray datasets}
\label{sec:dataset_xray}

\pks has been regularly observed by the X-ray observatories \rxte, \swiftxrt\  and \xmm. Preliminarily, we have gathered both monitoring and target of opportunity (ToO) observations of \pksc. However, since ToO observations bias the dataset towards high flux values, only monitoring data have been considered. Further, we applied a correction for the Galactic absorption with $n_{\rm H} = 1.48 \times 10^{20}$ cm$^{-2}$, according to~\citet{2005A&A...440..775K}.

\rxte\ data consist in publicly available\footnote{\url{http://cass.ucsd.edu/rxteagn/}} pre-analyzed light curves in four energy ranges: the full range from 2 to 10\,keV and three subranges 2--4, 4--7 and 7--10\,keV.

\swiftxrt\  data in the energy bands 0.3--2, 2--4, 4--7, 7--10 and 2--10\,keV have been analyzed using the package \texttt{HEASOFT 6.16}. The data were recalibrated using the last update of \texttt{CALDB} and reduced using the standard procedures \texttt{xrtpipeline} and \texttt{xrtproducts}. Caution has been taken to properly account for pile-up effects for corresponding affected exposures, and spectral fits were performed using \texttt{Xspec 12.8.2} assuming a power-law spectrum.

\xmm\  public EPIC (European Photon Imaging Camera) data in the energy ranges 0.3--2, 2--4, 4--7, 7--10 and 2--10\,keV have been reduced using the \texttt{SAS} software package (version 14.0) and analyzed following \citet{2012A&A...546A..88T}.

\begin{figure*}
\centering
\includegraphics[width=0.95\textwidth]{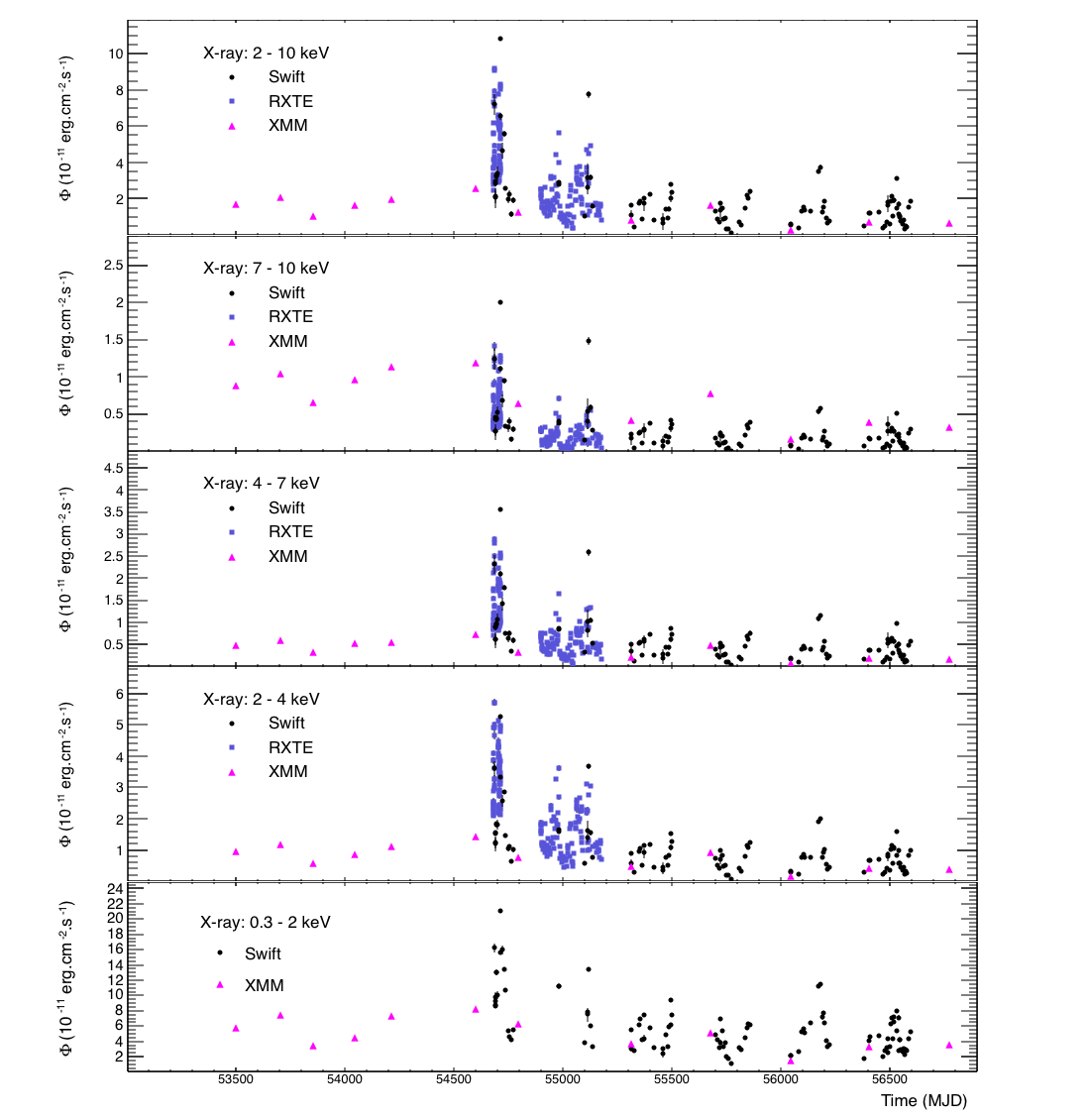}
\caption{X-ray light curves in the different energy ranges presented in Section~\ref{sec:dataset_xray}. Black points: \swiftxrt, pink triangles: \xmm, blue squares: \rxte. From bottom to top: 0.3--2\,keV, 2--4\,keV, 4--7\,keV, 7--10\,keV and the full common range 2--10\,keV.}
\label{fig:xray_lc}
\end{figure*}

Figure~\ref{fig:xray_lc} presents the X-ray light curves in the total 2-10 keV range as well as in the 4 energy sub-ranges.

\subsection{Optical dataset}

\begin{figure*}
\centering
\includegraphics[width=0.95\textwidth]{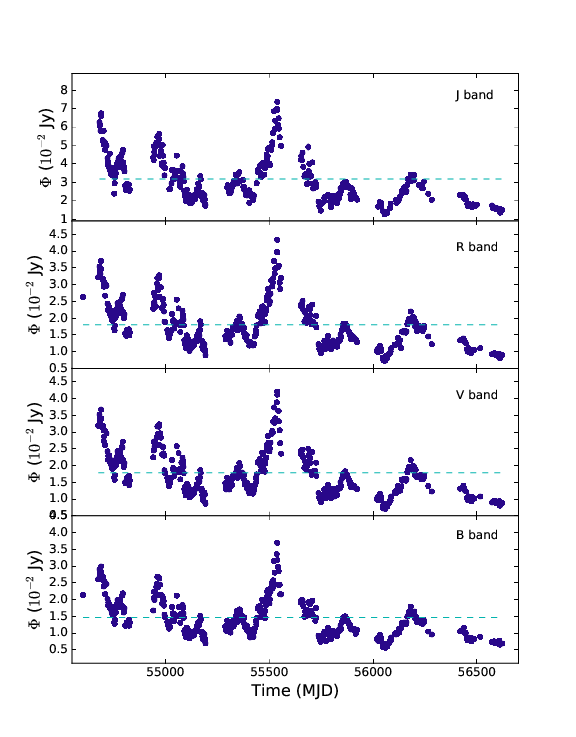}
\caption{SMARTS light curves in the different available bands (from top to bottom): J, R, V and B.}
\label{fig:smarts_lc}
\end{figure*}

SMARTS \citep[Small and Moderate Aperture Research Telescope System,][]{2012ApJ...756...13B} data are publicly available\footnote{\url{http://www.astro.yale.edu/smarts/glast/home.php}}. Magnitudes have been corrected for the absorption of the Galactic foreground following~\citet[]{2011ApJ...737..103S} and converted to a spectral flux density using the zero flux values of \citet[]{1992AJ....104.2030C}. The light curves, shown in Fig.~\ref{fig:smarts_lc}, are taken in four bands (J, R, V and B) in the same time windows as \fer\ (MJD 54603 to 56622).

\section{Results}
\label{sec:results}
\subsection{Fractional variability}

\begin{figure*}
\centering
\includegraphics[width=0.95\textwidth]{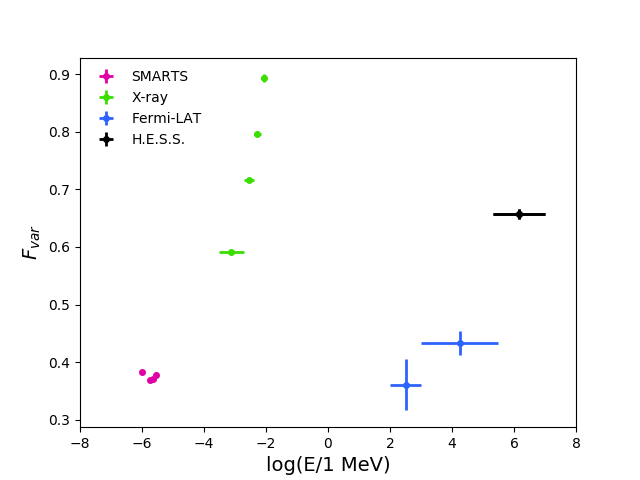}
\caption{Variability energy distribution of \pksc. SMARTS, \rxte, \swiftxrt\  and \xmm\ data were analysis in this work. \fla\ and \hess\ data were extracted from \citet{2015arXiv150903104C} and \citet{2016arXiv161003311H}, respectively.}
\label{fig:fvardata}
\end{figure*}

To study the variability of \pks, the fractional variability $\fvar$ as defined in~\citet{2003MNRAS.345.1271V} has been computed for each energy bin (Table \ref{table:fvar}). Figure~\ref{fig:fvardata} presents the evolution of $\fvar$ as a function of the energy (hereafter variability energy distribution). The value in the TeV range computed in~\citet{2016arXiv161003311H} is also reported. There is a clear trend, with  $\fvar$ increasing with energy from the optical range to X-ray. In \grs, $\fvar$ is lower than in X-ray but also increases with energy. This bimodal structure has been already reported for this source in \citet{2009ApJ...696L.150A} but is shared also by other objects, e.g. Mrk421 \citep{2007A&A...462...29G,2016A&A...593A..91A}.

\begin{table*}
\centering
\caption{$\fvar$ values for each energy range of the \pks\ data set presented in this work.}
\label{table:fvar}
\begin{tabular}{@{}cc@{}}
\hline
 Energy range & $F_{\rm var}$ \\
\hline
0.2--10\,TeV$^a$	& $0.657 \pm 0.008$ \\
0.1--1\,GeV$^a$	& $0.36 \pm 0.04$ \\
1--300\,GeV$^a$			& $0.43 \pm 0.02$ \\
0.3--2\,keV 			& $0.591 \pm 0.004$ \\
2--4\,keV  				& $0.716 \pm 0.003$ \\
4--7\,keV 				& $0.796 \pm 0.004$ \\
J 							&  $0.383 \pm 0.005$ \\
R 							&  $0.369 \pm 0.003$ \\
V 							&  $0.371 \pm 0.004$ \\
B 							& $0.378 \pm 0.004$ \\
\hline
\multicolumn{2}{l}{$^a$From \citet{2016arXiv161003311H}}\\
\multicolumn{2}{l}{$^b$From \citet{2015arXiv150903104C}}
\end{tabular}
\end{table*}

\subsection{Log-normality of the flux}

We fit the X-ray and optical flux distribution $\Phi$ and its logarithm $\log(\Phi)$ with a Gaussian. The results are summarized in Table \ref{table:excess}, also reporting the $\chi^2$ values. All light curves present a $> 3 \sigma$ preference for a log-normal distribution, i.e. the distribution of $\log(\Phi)$ is better described by a Gaussian than the distribution of $\Phi$.

\begin{figure*}
\centering
\includegraphics[width=0.49 \textwidth]{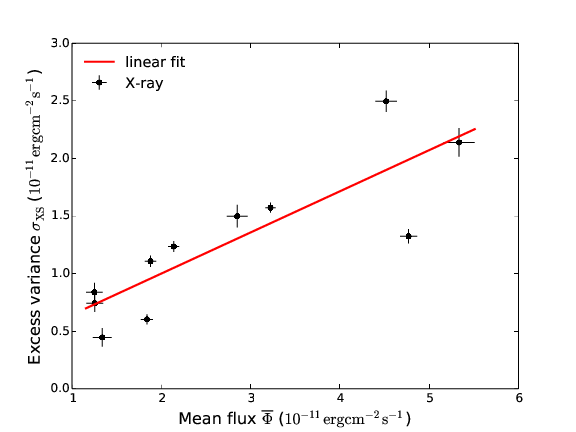}
\includegraphics[width=0.49\textwidth]{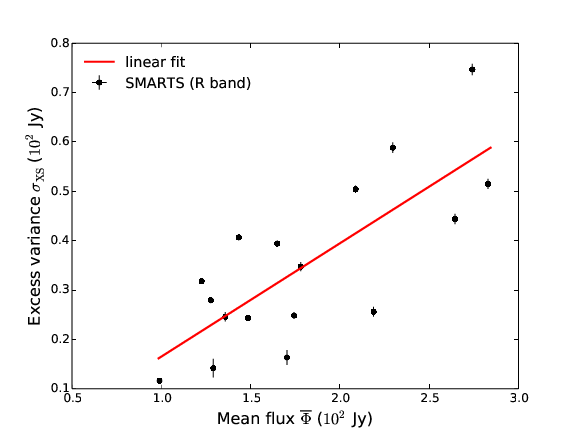}
\caption{Excess  variance $\sigma_{\rm XS}$ as a function of the mean flux $\overline{\Phi}$ in X-ray and in the SMARTS R band. The red line is the result of a linear fit to the data.}
\label{Fig:excessCor}
\end{figure*}

The excess variance $\sigma_{\rm XS}$ as defined in~\citet{2003MNRAS.345.1271V}  vs. the average flux $\overline{\Phi}$ is shown in Figure~\ref{Fig:excessCor}. Each light curve is split in several bins with at least 20 points to ensure sufficient statistics to compute $\sigma_{\rm XS}$ and $\overline{\Phi}$. These two quantities are clearly correlated,  and  linear fit is found to better reproduce the data than a constant fit (see Table \ref{table:excess}).
Although insufficient statistics has prevented to reach firm conclusions in the HE range, \citet{2016arXiv161003311H} has reported a similar behavior in the VHE range. Overall, these results are suggestive of multiplicative processes as main responsible for the variability of \pks through the whole spectrum. In such a stochastic process, a high flux leads to an increased variability which, in turn, possibly leads to higher flux.

Evidence for log-normality on different time scales has been reported for different sources: BL~Lacertae in X-ray \citep{2009A&A...503..797G}, in the VHE for the BL Lac Mrk5~01 \citep{2010A&A...524A..48T,2015arXiv150904893C} or for \pks, during flaring event in VHE \citep{2009A&A...502..749A}. This behavior was also observed for non-blazar objects, such as for the Seyfert 1 galaxy  IRAS~13244$-$3809 in X-rays \citep{2004ApJ...612L..21G}. There are growing evidences that this behavior is a common feature of accreting systems. In the context of galactic X-ray binaries, where log-normal flux variability has first been established, such a behavior is thought to be linked to the underlying accretion process \citep{2001MNRAS.323L..26U}. The detection of log-normality in a Seyfert 1 galaxy, a class of radio quiet AGN lacking a relativistic jet and whose emission line emission correlates with the amount of gas surrounding the central engine, strengthens the link between accretion disk and log-normal behavior.

\begin{table*}
\centering
\caption{Left: Values of $\chi^2$ for the Gaussian fit of $\Phi$ and $\log(\Phi)$, with values for the significance $\sigma$. Right: Values of the reduced $\chi^2$ of the constant and linear fits of the scatter plots shown in Fig.~\ref{Fig:excessCor} for each light curve. $\rho$ is the correlation factor. The corresponding data set is named in the first column.}
\label{table:excess}

\begin{tabular}{l c c c | c c c}
\hline \hline
& $\Phi$ & $\log(\Phi)$ & & constant & linear increase &  \\
& $\chi^2/$d.o.f. & $\chi^2/$d.o.f. & $\sigma$& $\chi^2/$d.o.f. & $\chi^2/$d.o.f.  & $\rho$  \\
\hline
X-ray 			& $80.0/12$ 		   & $12.5/9$  &  $7.69$	& $782/10$   & $260/9$    & $0.85 \pm 0.02$ \\
SMARTS (J) 		& $56.6/13$    		   & $5.1/12$  &  $7.18$ 	& $3077/13$  & $865/12$   & $0.81 \pm 0.01$ \\
SMARTS (R) 		& $29.9/13$            & $8.7/11$  &  $4.22$	& $22462/16$ & $7858/15$  & $0.93 \pm 0.02$ \\
SMARTS (V) 		& $65.1/12$  		   & $9.1/11$  &  $7.48$	& $3800/15$  & $1746/14$  & $0.76 \pm 0.01$ \\
SMARTS (B) 		& $30.2/13$            & $15.4/12$ &  $3.85$	& $3676/15$  & $2234/14$  & $0.72 \pm 0.01$ \\
\hline \hline
\end{tabular}
\end{table*}

\subsection{Search for periodicity}

 \citet{2014ApJ...793L...1S} reported a possible periodic behavior in the optical and HE light curves of \pksc \ and an intriguing coincidence of a period in HE roughly double the one in optical was noted.
  To study further such periodic features, the multi-wavelength light curves are analyzed with a Lomb Scargle periodogram \citep[LSP;][]{1976Ap&SS..39..447L,1982ApJ...263..835S}.
The LSP is a method to estimate the Power Spectrum Density (PSD) of a time series based on a least squares fit of sinusoids to the data sample. The advantage of the LSP compared to a classical Fourier analysis is that it takes into account the uneven spacing of the data. The standard LSP was used \citep[section 5 of ][]{2018ApJS..236...16V}, as implemented in the {\tt astropy} package \citep{2018AJ....156..123A}.

In this work, the light curves are not evenly sampled and binned differently. Moreover, gaps between observations - appearing from the impossibility to observe the source during some period of the year - have to be taken into account. For all the analyses, the maximum frequency 
is set following section 4.1.3. of \citet{2018ApJS..236...16V} to $f_\mathrm{min}=1/(2\delta t)$ where $\delta t$ is the typical integration time. To have uniform results, this has been set according to the most constraining data set, i.e. the \fla\ integration time (10 days).

\begin{figure*}
\centering
\includegraphics[width=0.47\textwidth]{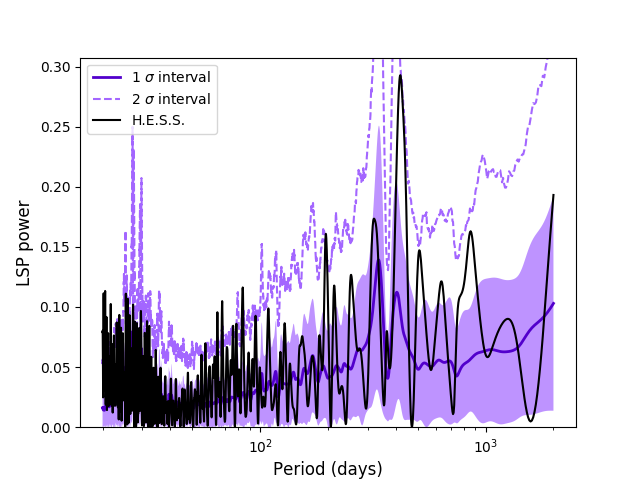}
\includegraphics[width=0.47\textwidth]{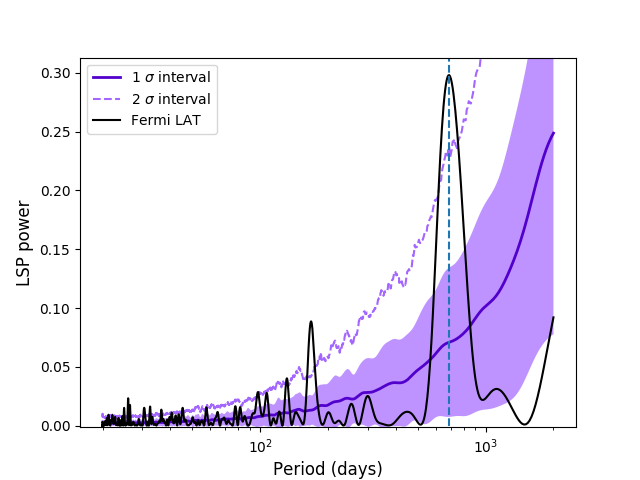}
\includegraphics[width=0.47\textwidth]{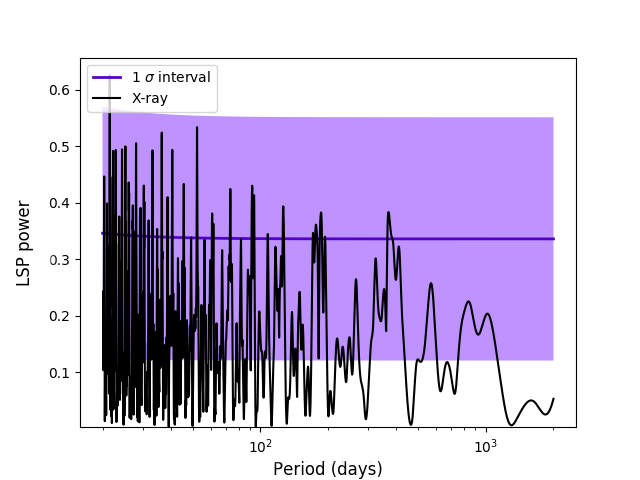}
\includegraphics[width=0.47\textwidth]{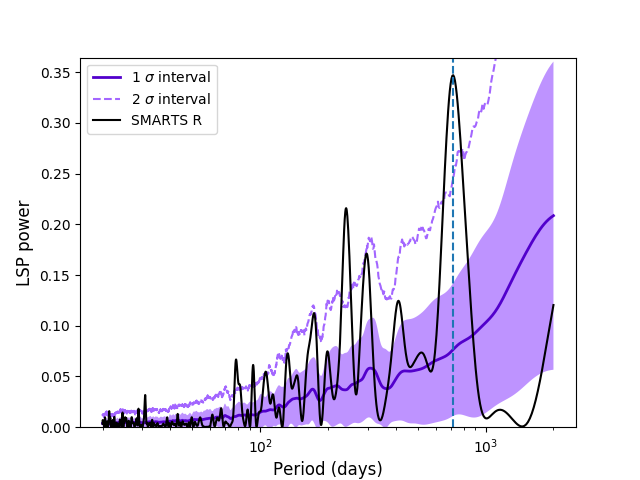}
\caption{Lomb Scargle periodogram for the R band of SMARTS (top left), the full range X-ray light curve (top right), the full range \fla\ data (bottom left) and the \hess\ full range (bottom right). The black line represents LSP applied to the data, while the purple area is the local 1 $\sigma$ confidence interval coming from the simulations, with the mean value represented by the solid purple line. The dashed purple line encloses the local 2 $\sigma$ confidence interval coming from the simulations. The vertical, dashed blue line marks the most prominent periodicity.}
\label{Fig:lsp}
\end{figure*}

The LSPs of the SMARTS (in the R band only for the sake of clarity), X-ray, \fla\ and \hess\ light curves are shown in Fig.~\ref{Fig:lsp} along with the 1$\sigma$ and 2 $\sigma$ confidence intervals. The X-ray and H.E.S.S. light curves do not show any periodic feature. In optical, the B,R,V and J bands exhibit a significant periodicity, with the best fit period ranging from $715$ to $733$ days depending on the band, while the HE light curve is found to have a periodicity of 685$\pm 9$ days.
 \citet{2014ApJ...793L...1S} found a similar period in the HE range as the one reported here, but a T$\approx$315 day period for the optical light curve. While we do confirm the presence of a peak in the LSP of optical data around 300 days, the most intriguing excess of the power is at $\simeq$700 days (see Fig.~\ref{Fig:lsp}), since it is found both in optical and HE light curves.

In order to assess the significance of our results, light curves without periodicity have been simulated and rebinned according to the observational time binning. This allows one to factor out instrumental effects such as windowing due to sparse observation and/or binning in time due to limited sensitivity.  Each simulated light curve has been obtained by inverse Fourier transform of power-law noise, without adding a constant term, adopting a different spectral index for each energy ranges. In VHE, an index of 1.1 has been taken while in HE, the used value is 1.2. \citep{2016arXiv161003311H}. In optical and X-ray, an index of 1.2 has been used. We have also checked that the results are robust with respect to slight ($\sim 10-20\%$) changes of the spectral index, representative of typical fit uncertainties.Then, for each period, the range of LSP power spanned by 68\% (95\%) of the realizations is used to derive the 1 $\sigma$ (2 $\sigma$) contours shown in Fig.~\ref{Fig:lsp}.

One limitation of the visual inspection of  Fig.~\ref{Fig:lsp} is that it cannot obviously account for the trial factor effect, coming from a scan over different frequencies tested. 
Note that the {\tt astropy} python package used here can provide an estimate of a false alarm probability (FAP) taking into account the trial factor; however, it 
implicitly assumes non-varying data with Gaussian noise, while the real data follow a red noise behavior. As such, this estimator cannot be taken at face value. We merely use it to perform some sanity checks, e.g. to verify that the FAP computation following \citet{2008MNRAS.385.1279B} yields more conservative results than the method~\citep{2018ApJS..236...16V}, as expected.

Despite this limitation, our results remain intriguing: While taken separately each of the peaks found in the right panels of Fig. 5 might not be statistically very significant,  by interpreting e.g. the \fla\ results as a test search to suggest the most interesting periods to search {\it a priori} in the optical bands, the  $\sim 3\,\sigma$ excess found in SMARTS data sample at comparable period of $\simeq$700 days can be taken more or less at face value, since (most of) the trial factor is basically accounted for in the \fla\ sample search. Albeit heuristic, this argument is also what singles out this period compared to others, for which no matching is seen in the multiwavelength comparison.

Besides the above-mentioned technical difficulties, these kinds of long-term periodicity analyses suffer from physical limitations, such as the fact that only a few periods are probed. It is also worth noting that in a recent article, \citet{2018arXiv181002409C} warned that for none of the blazars whose periodicity in \fla\  band is reported in \citep[]{2017MNRAS.471.3036P,2017A&A...600A.132S,2017ApJ...842...10Z,2017ApJ...845...82Z} a strong statistical case can be made.
However, \citet{2018arXiv181002409C} only used  \fer\ data; one comforting cross-check from the multiwavelength data presented here is that cross-correlations between LAT and optical data are significant both at times $\tau=0$ (no delay) and at a timescale approximately equal to the reported periodicity. We take the {\it simultaneous} hints for a periodicity around 700 days in both \fer\ and optical datasets as the most intriguing indication of our analysis. While awaiting a definitive confirmation in {\it a priori} searches in future independent datasets, in the next section we will tentatively considering the implications of including or not this periodicity for the interpretations in the context of a simple SSC model.

\section{Time-dependent modeling}
\label{sec:sscmodeling}
\subsection{The synchrotron self-Compton model and the steady-state of PKS~2155-304}
\label{sec:steadystate}

In order to reproduce the results found in this work and especially the variability evolution with energy, a one-zone synchrotron SSC model has been considered \citep{THEO::SSC_BAND}. In this model, the first bump of the SED is produced by the synchrotron radiation of electrons spinning into the uniform magnetic field $B$ of the jet. The second bump of the SED is explained by inverse Compton scattering of the same electrons population on the previous synchrotron photon field. The emission is assumed to be produced by a homogeneous region of radius $R$ propagating in the jet with a Doppler factor $\delta$.

If electrons are injected with a time dependent function $Q(E,t)$ and radiate their energy via synchrotron or inverse Compton processes, the electron density $N_e$ is given by the diffusion equation:
$${{\rm d} N_e(E,t) \over {\rm d} t } = {\partial \over \partial E} \left[\left(\dot{\gamma_{\rm S}}+\dot{\gamma_{\rm IC}}\right)N_e(E,t)\right] + Q(E,t),$$
where $\dot{\gamma_{\rm S}}$ and $\dot{\gamma_{\rm IC}}$ are the synchrotron and inverse Compton cooling rates of the electrons. The escape of the electrons is not taken into account in this model. If the escape time scale is larger than the cooling time scale, this has no effect in the model. Lower time scales will lead to an achromatic decrease of the variability. The injection of the electrons $Q(E,t)$ is chosen to be a power-law with exponential cut-off:
$$Q(E,t) = N_0(t) E^{-\alpha(t)} \exp\left({- E \over \gamma_{\rm cut}(t)}\right),$$
where $N_0$ is the injection normalization, $\alpha$ the power-law index and $\gamma_{\rm cut}$ the energy of the exponential cut-off.

The equation is solved numerically using the algorithm of \citet{CC70} for each time step, which allows us to follow the evolution of the electron density and hence of the emitted flux. The time-averaged SED has been modeled by reaching the steady state of the diffusion equation for $Q(E,t) \equiv Q(E)$. The parameters used are given in Table~\ref{table:steadystate}.

\begin{table*}
\centering
\caption{SSC model parameters of the steady state of \pksc.}
\label{table:steadystate}
\begin{tabular}{l c c}
\hline \hline
Normalisation & $N_0$ & $2.7 \times 10^{47}$ electrons \\
Index & $\alpha$ & $2.3$ \\
Cut off energy & $\log(\gamma_{\rm cut} / 1 {\rm~eV})$ & $5.3 $ \\
Magnetic field & $B$ & $0.1$ G \\
Radius & $R$ & $2 \times 10^{16}$ cm \\
Doppler factor & $\delta$ & 35 \\
\hline \hline
\end{tabular}
\end{table*}

\subsection{Simulation of the variability}

To introduce the variability in the model used in this work, one of the parameters of the model was chosen to vary with time. \citet{2007A&A...462...29G} modeled the emission of Mrk~421 with a similar model and an injection function being a relativistic Maxwellian function. They found that two flux states (high flux and low flux) can be reproduced by merely changing the characteristic energy. Most interesting is that they predicted small flux variations in optical and GeV ranges and high variations in X-ray and TeV range. In the modeling presented here, $\gamma_{\rm cut}$ is similar to their Maxwellian characteristic energy. Since the number of injected particles increases exponentially with $\gamma_{\rm cut}$, the relevant parameter is then $\log(\gamma_{\rm cut})$.

AGN light curves generally show power-law noise of the form $1/f^\beta$ \citep{1993ApJ...414L..85L}. To reproduce this property, $\log(\gamma_{\rm cut})$ is varied during the simulations following a power-law noise of index $\beta$ and total variance $\sigma$, and a mean of $\log(\gamma_{\rm cut})=5.3$. Simulations of the variation of $\log \gamma_{\rm cut}$ are drawn following \citet{1995A&A...300..707T}.  Note that time series of $\log \gamma_{\rm cut}$ were constructed on a timescale ten times longer than the needed amount, to ensure that long term variations are well reproduced using this technique.

Simulations with $\beta \in$ [1.0, 1.5, 2.0] and $\sigma \in$ [10\%, 15\%, 20\%, 25\%] were performed to find the couple that best reproduces the variability energy distribution. In total, 200 simulations of 10 years each with a binning of 9.5 minutes in the observer rest frame (to ensure that small variations are simulated) were performed.  Values of $\beta = 1.0$ (i.e. flicker noise) and $\sigma $= 20\% were found to best match the variability energy distribution.  The results of the simulations are found to be mostly sensitive to $\sigma $. Indeed, increasing this parameters increases the measured variability mainly in optical and \fer\ energy ranges.

\begin{figure*}
\centering
\includegraphics[width=0.95\textwidth]{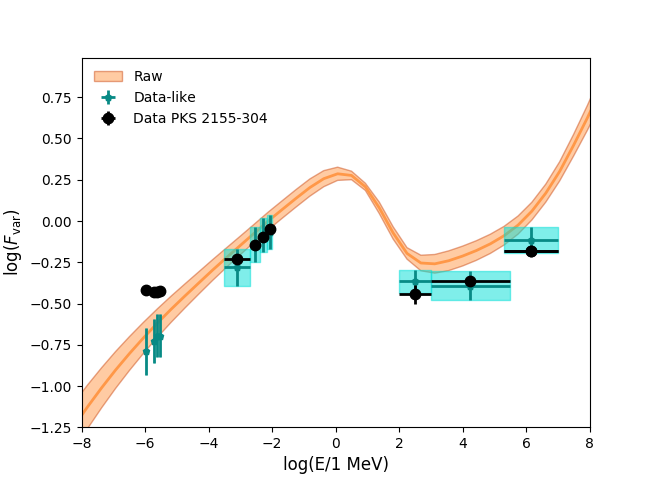}
\caption{Variability energy distribution $F_{\rm var}(E)$ for the best configuration of the power-law noise of $\log(\gamma_{\rm cut}(t))$ with $\beta = 1$ and $\sigma = 20\%$. The black points are the $F_{\rm var}$ of the data presented in Sec.~\ref{sec:dataset}. The orange curve is the $F_{\rm var}(E)$ for the simulated light curves while the cyan points represent the $F_{\rm var}$ for the simulated light curves rebinned in energy and in time.}
\label{fig:fvar_result}
\end{figure*}

Figure~\ref{fig:fvar_result} shows again the variability energy distribution (black points as from Table~\ref{table:fvar}) but also reports the simulated $\fvar$ as an orange band. The variability increases from the lowest energy up to the X-ray domain, then drops in the HE range, and eventually increases towards TeV energies, following the same trend as the data, albeit quantitative discrepancies are noticeable.

To take into account the differences between each instrument, each simulated light curve is rebinned in time and energy following the observations of each instrument.
The cyan points/boxes represent the variability of the simulated light curves after rebinning.  Such binning does not change much the pattern in optical, X-ray and TeV energies. However, the variability in the \fla\ ranges is sensibly reduced, now matching the data within errors. This is likely due to the important time bins (10 days) used for the \fla\ analysis. The only band which appears in clear disagreement with the model is the optical range, where the variability of the SMARTS data is not reproduced; another source of variability has to be invoked.

\subsection{Power spectral density}
\label{sec:psd}

\begin{figure*}
\centering
\includegraphics[width=0.49\textwidth]{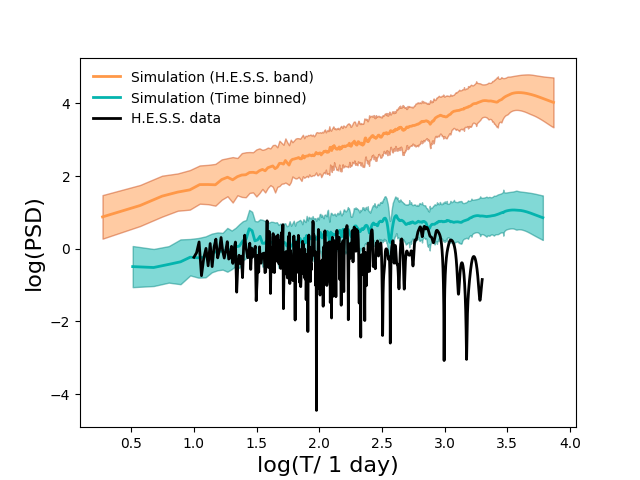}
\includegraphics[width=0.49\textwidth]{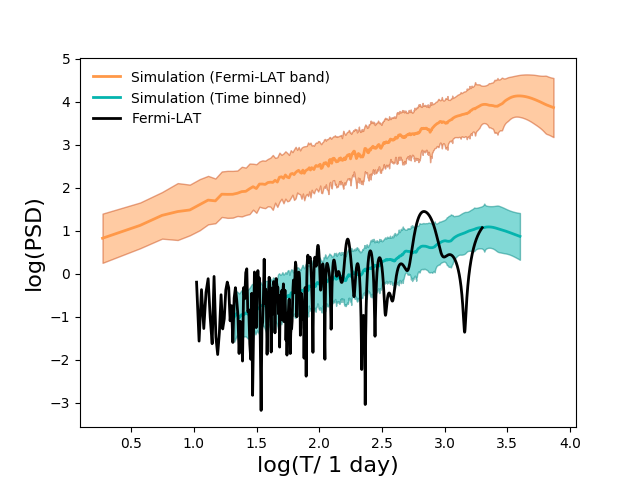}
\includegraphics[width=0.49\textwidth]{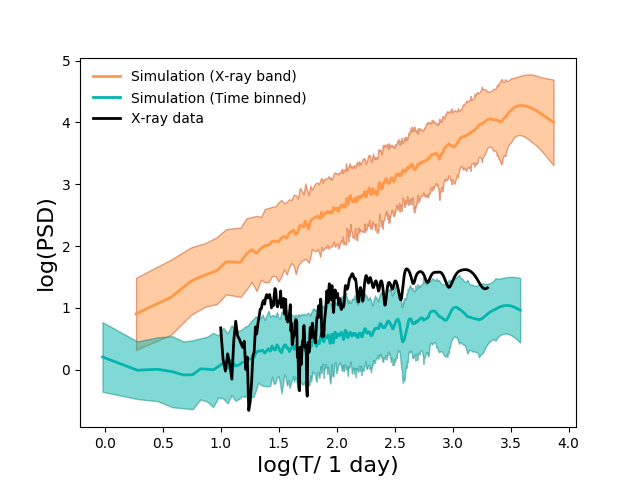}
\includegraphics[width=0.49\textwidth]{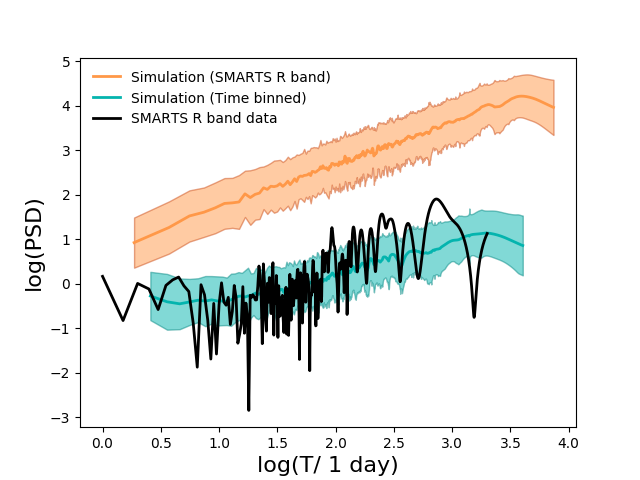}
\caption{PSD of the H.E.S.S., \textit{Fermi}-LAT, X-ray and SMARTS energy ranges (from top to bottom). The black curve is the PSD of the data. The orange curve represents the PSD of the simulated light curves with no time-binning applied while the blue one is for the rebinned simulated light curves.}
\label{fig:psd_result}
\end{figure*}

In~\citet{2016arXiv161003311H}, both the HE and VHE power spectral density  were computed. It was found that they can be  quite well characterized by a flicker noise, the best-fit power-law indexes being
 $\beta_{\rm HE} = 1.20^{+0.21}_{-0.23}$ and $\beta_{\rm VHE} = 1.10^{+0.10}_{-0.13}$, respectively. For completeness,  we report in Figure~\ref{fig:psd_result} a comparison of PSD obtained in  the simulations (orange-shaded areas) vs. the multiwavelength data (VHE, HE, X-ray and optical/R band).  The cyan band accounts for the correct binning in time, according to the actual observations: notice how this effect is crucial for the simulation to match the data. It is noticeable that this is an important consistency check, since nothing in the simulation was tuned to reproduce the PSD.

\subsection{Adding a periodic component}
The hinted periodicity of \pks in the optical band (and, to some extent, at HE) could be explained by different physical effects.

The most frequently discussed culprit for quasi-periodic behavior in blazars are quasi-periodic modifications of the Doppler effect.
A fascinating possibility is that a binary supermassive black hole (SMBH) system could be at the center of these AGN, instead of just one SMBH as assumed in the general AGN picture \citep{1980Natur.287..307B}. This binary system could cause a periodic change in the accretion rate of the matter coming from the disk and even misalign the accretion disk \citep[][and references therein]{2015MNRAS.449.1251D}. A similar outcome may be due to the Lense-Thirring effect, breaking the central regions of tilted accretion disks around spinning black holes, see e.g.~\citep[][, and references therein]{2015MNRAS.449.1251D}.
These scenarios however face the difficulty that jet precession is expected to happen on too long time scales, $\sim 10^6$~years according to the analyses of~\cite{2008MNRAS.385.1621K} and \cite{2013ApJ...765L...7N}.
A recent study~\citep{Sandrinelli:2018rdl} also points out the tension that a binary SMBH origin associated to (the relatively common) blazar periodicity may have with pulsar timing array limits on the gravitational wave emission of such close SMBH binaries.
In~\citet{Raiteri:2017cul}, optical-to-radio monitoring of the blazar CTA~102 has been argued to support a scenario where magnetohydrodynamic instabilities or the rotation of a twisted jet cause different jet regions to change their orientation\footnote{however see e.g. \citet{2017ApJ...851...72Z} for an alternative interpretation of the recent variability exhibited in CTA~102.}, hence their relative Doppler factors. Other observational evidence in the AGN BL~Lacertae and M~81 suggesting a precession motion of their jets looking at radio knots with VLBI observations can be found in~\citet{2003MNRAS.341..405S,2013MNRAS.428..280C,2013arXiv1301.4782M}.

With the aim to test if a periodic variation of the Doppler factor can account for the observations of \pks, we performed simulations analogous to what previously described, but adding on top of the stochastic variation of $\log(\gamma_{\rm cut})$ a sinusoidal time series $\delta (t) = \delta_\textnormal{steady state} + 5 \times \sin(t + T ) $. We fix $\delta_\textnormal{steady state} = 35$ from what is found in Section~\ref{sec:steadystate}. The amplitude is an had-oc value.

This addition yields an increase of the global variability in an achromatic way. To compensate for this effect, the simulations were redone with $\sigma_{\rm cut} = 15 \%$, keeping $\beta=1$.

\begin{figure*}
\centering
\includegraphics[width=0.95\textwidth]{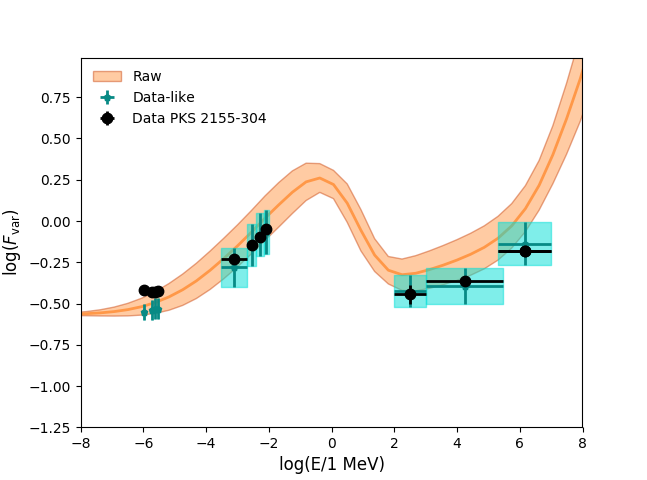}
\caption{Same as Figure~\ref{fig:fvar_result} but with periodic change in the Doppler factor and with  $\sigma = 15\%$.}
\label{fig:fvar_result_per}
\end{figure*}

 The resulting variability energy distribution is shown in Figure~\ref{fig:fvar_result_per}. The shape of the variability energy profile stays roughly the same, with however a flatter part in the optical range around $F_{\rm var} = 0.20-0.25$, rising the variability levels of the simulation in this range compared to the non periodic one (Figure~\ref{fig:fvar_result}).

It is clear that the periodicity can (at least partially) explain why the SMARTS data are more variable than the previously considered model. Within this new scenario, this energy range would be dominated by the variability of the periodicity and not by the one of the stochastic process. This is not surprising, since a modification of the value of the cut-off energy has a small impact on the electrons producing the optical photons. It is also encouraging that a minor deviation between low-energy \textit{Fermi} data and our simulations present in Figure~\ref{fig:fvar_result} shrinks to an irrelevant difference in presence of periodicity.

\begin{figure*}
\centering
\includegraphics[width=0.47\textwidth]{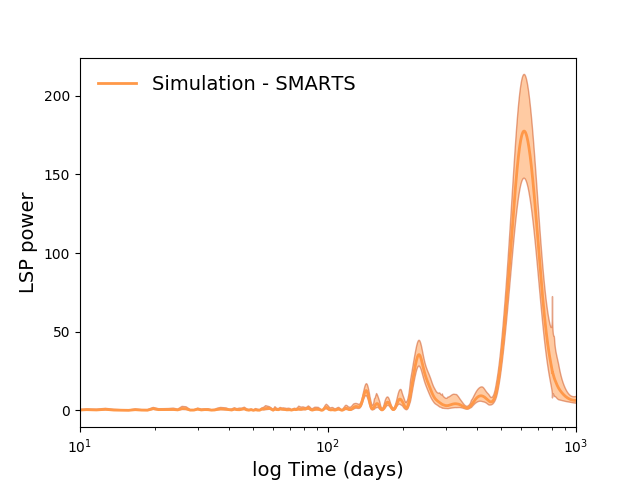}
\includegraphics[width=0.47\textwidth]{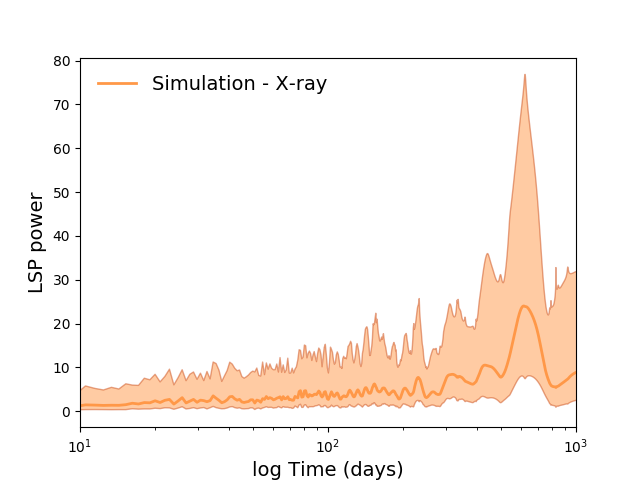}

\includegraphics[width=0.47\textwidth]{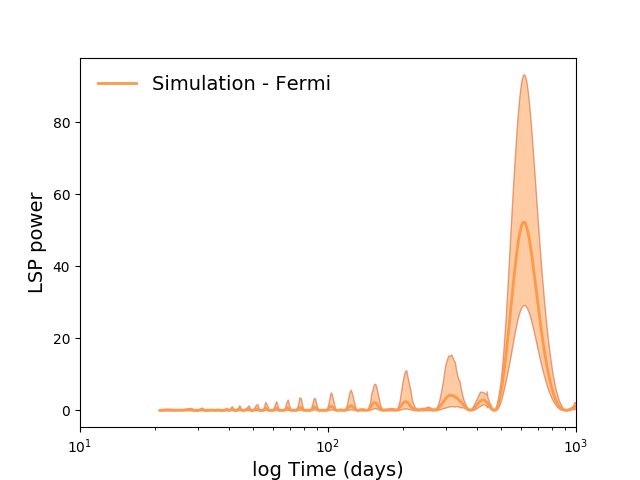}
\includegraphics[width=0.47\textwidth]{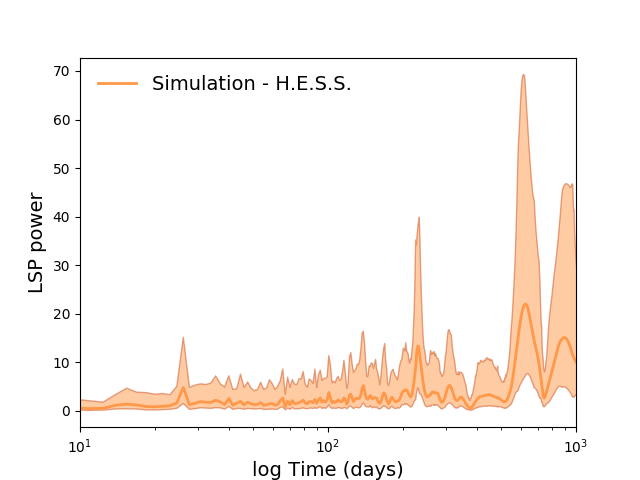}
\caption{LSP of the simulated light curves with a periodicity of $600$ days injected in the Doppler factor for the optical R, X-ray, \textit{Fermi}-LAT and H.E.S.S. bands (from top to bottom).}
\label{Fig:lsp_ssc_deltaper}
\end{figure*}

Figure~\ref{Fig:lsp_ssc_deltaper} displays the periodograms of the simulated light curves in the SMARTS, X-ray, \textit{Fermi}-LAT and H.E.S.S. ranges, without applying any temporal binning on the short term simulated light curves. The optical and GeV simulated light curves have a clear and strong bump around 600-700~days, being appreciably (if not dominantly, for the optical) sensitive
to the periodic variation of $\delta$,  while the X-ray and TeV light curves show a less significant bump, since they are especially sensitive to the stochastic change of $\gamma_{\rm cut}(t)$. This is qualitatively consistent with the lack of detection of periodicity in X-ray or VHE ranges till now.

\section{Discussion and conclusions}\label{sec:discconcl}

Multiwavelength data spanning up to 10~years of observations of the blazar \pks\ have been gathered and studied in this work. SMARTS optical, \rxte, \swiftxrt\ and \xmm\,have been analyzed, as well as  HE and VHE gamma-ray data taken from \fer\ and \citet{2016arXiv161003311H}, respectively. Optical and HE ranges shows only little variability with respect to the variability found in X-ray or in VHE. This seems to be a characteristic shared with other BL Lac (e.g. Mrk~421) and indicate a close link in the population of particles that emits the low and high energy parts of the SED.

As in VHE \citep{2016arXiv161003311H}, X-ray and optical bands exhibit a log-normal behavior, indicative of a multiplicative process. This is also an argument for a link in the origin of the variability of both bands. Nevertheless, the tentative indication for periodicity around 700 days found in optical and HE is observed neither in X-ray nor in VHE.

The time-dependent SSC model used in this work explains well the evolution of the variability with energy, except for the optical band. Adding a periodic component in the model helps in better describing this energy range, and also improves the agreement in the HE range. Although model-dependent, this is another encouraging indication that stimulates further studies to confirm the hint of a periodicity reported here with future, independent datasets.
Our model also reproduces the non-detection of the periodic behavior in X-ray and VHE bands. Independently of how realistic the models discussed are, our results are a healthy reminder that depending on the energy range of interest, the mechanism(s) dominating the observed variability can be different.

Still, some questions remain on the origin of the variability of the whole spectrum of \pksc. For instance, it is worth keeping in mind that the observed log-normal behavior is not explained in the models discussed in this article, and a consensual quantitative theory of its microscopic origin is still lacking. Definitely, more long-term observations of different blazars would be needed to extract common features and differences between objects, in turn helping refining the theoretical models.


\section*{Acknowledgements}

This work has been done thanks to the facilities offered by the Université Savoie Mont Blanc MUST computing center, as well as by CC-IN2P3 (\href{https://cc.in2p3.fr}{cc.in2p3.fr}).

This research has made use of NASA's Astrophysics Data System, as well as data and/or software provided by the High Energy Astrophysics Science Archive Research Center (HEASARC), which is a service of the Astrophysics Science Division at NASA/GSFC and the High Energy Astrophysics Division of the Smithsonian Astrophysical Observatory.


\bibliographystyle{mnras}
\bibliography{LT_PKS2155}

\end{document}